\documentclass[utf8]{frontiersSCNS} 

\usepackage{url,hyperref,lineno,microtype,subcaption}
\usepackage[onehalfspacing]{setspace}
\usepackage{natbib}

\def\keyFont{\fontsize{8}{11}\helveticabold }
\def\firstAuthorLast{Mendoza {et~al.}} 

\def\Authors{Edgar Mendoza,$^{1,3}$  Nicolas Duronea,$^2$ Daniele Rons\'o,$^3$ Lia C. Corazza,$^4$\\
Floris van der Tak,$^{5,6}$ Sergio Paron,$^7$ Lars-\AA ke Nyman$^8$}

\def\Address{$^1$Observatório do Valongo, UFRJ, Rio de Janeiro, Brazil\\
$^2$Instituto de Astrof\'isica de La Plata, La Plata, Argentina\\ 
$^3$Instituto de Astronomia, Geof\'isica e Ci\^encias Atmosf\'ericas, USP, S\~ao Paulo, Brazil\\
$^4$Instituto Nacional de Pesquisas Espaciais, S\~ao Jos\'e dos Campos, Brazil\\
$^5$SRON Netherlands Institute for Space Research\\
$^6$Kapteyn Astronomical Institute, University of Groningen, The Netherlands\\
$^7$CONICET-Universidad de Buenos Aires. Instituto de Astronomía y Física del Espacio, Buenos Aires, Argentina\\
$^8$European Southern Observatory, Alonso de Córdova 3107, Vitacura, Santiago, Chile
}

\def\corrAuthor{Edgar Mendoza}

\def\corrEmail{emendoza@astro.ufrj.br}

\begin{document}
\onecolumn
\firstpage{1}

\title[Astrochemistry \& Galactic Dynamics]{Interrelations between Astrochemistry and Galactic Dynamics} 

\author[\firstAuthorLast ]{\Authors} 
\address{} 
\correspondance{} 

\extraAuth{}

\maketitle 

\begin{abstract}

This paper presents a review of ideas that interconnect Astrochemistry and Galactic Dynamics. Since these two areas are vast and not recent, each one has already been covered separately by several reviews. After a general historical introduction, and a needed quick review of processes like the stellar nucleosynthesis which gives the base to understand the interstellar formation of simple chemical compounds (e.g. H$_2$, CO, NH$_3$ and H$_2$O), we focus on a number of topics which are at the crossing of the two big areas, Dynamics and Astrochemistry. 

Astrochemistry is a flourishing field which intends to study the presence and formation of molecules as well as the influence of them into the structure, evolution and dynamics of astronomical objects.  The progress in the knowledge on the existence of new complex  molecules and of their process of formation originates from the observational, experimental and theoretical areas which compose the field. The interfacing areas include star formation, protoplanetary disks, the role of the spiral arms and the chemical abundance gradients in the galactic disk. It often happens that the physical   conditions in some regions of the interstellar medium are only revealed by means of molecular observations.

To organise a rough classification of chemical evolution processes, we discuss about how astrochemistry can act in three different contexts: namely, $i.$ the chemistry of the early universe, including external galaxies, $ii.$ star forming regions, and $iii.$ asymptotic giant branch (AGB) stars and circumstellar envelopes.  We mention that our research is stimulated by plans for instruments and projects, such as the on-going Large Latin American Millimeter Array (LLAMA), which consists in the construction of a 12m sub-mm radio telescope in the Andes. Thus, modern and new facilities can play a key role in new discoveries not only in astrochemistry but also in radio astronomy and related areas. Furthermore, the research of the origin of life is also a stimulating perspective.

\tiny
 \keyFont{ \section{Keywords:} Astrochemistry, galaxies: general, ISM: molecules, methods: miscellaneous, history and philosophy of astronomy.} 
 
\end{abstract}

\section{Introduction}

\subsection{Early impressions on the Milky Way}

\begin{figure}[h!]
\begin{center}
\includegraphics[width=16cm]{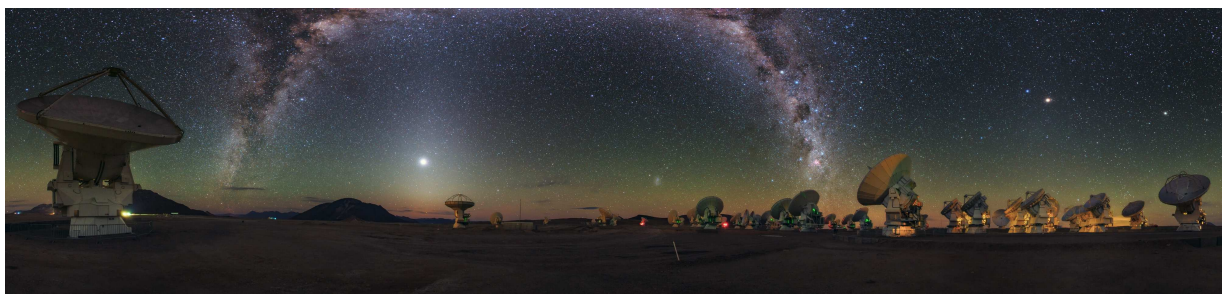}
\end{center}
\caption{Panoramic image capturing the Milky Way in a night sky arching over the ALMA Observatory's antennas in
the Chajnantor plateau (Chilean Andes). A cosmic rainbow in ultra HD, credit:
ESO/B. Tafreshi (twanight.org).}\label{fig1}
\end{figure}

\begin{figure}
\begin{center}
\includegraphics[width=15cm]{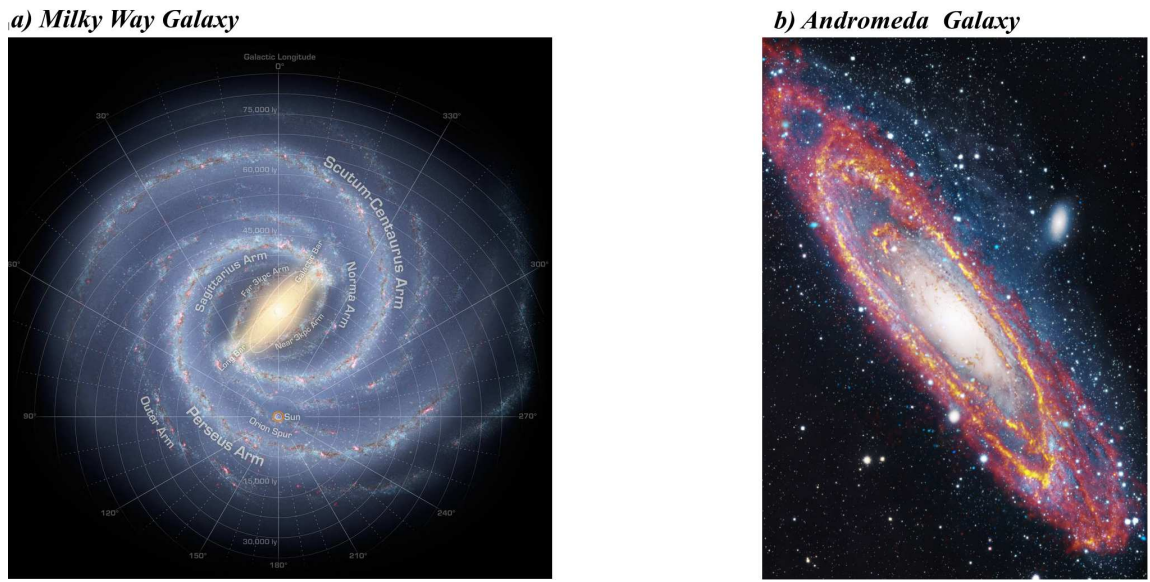}
\end{center}
\caption{a) Artistic view of the Milky Way Galaxy (diameter $\sim$ 100 klyr) showing its arms and central bar with respect to the Sun's position, which lies near the so-called Orion Arm, or Orion Spur, between the Sagittarius and Perseus arms (Credit: NASA/JPL-Caltech/R. Hurt, SSC/Caltech). b) The Andromeda Galaxy (diameter $\sim$ 220 klyr) in a composite image from three parts of the electromagnetic spectrum, the infrared, X-ray and optical (Credit: ESA/Herschel/PACS/SPIRE/J.Fritz, U.Gent/XMM-Newton/EPIC/W. Pietsch, MPE/R. Gendler). The Milky Way and Andromeda  are neighbouring galaxies expected to collide in a few billion years forming a merge  galaxy (\lq\lq Milkomeda\rq\rq)  \citep{Schiavi2020}.}\label{fig2}
\end{figure}

One of the most fundamental questions of mankind is why there is something rather than nothing \citep{Stavinschi2011,Allen2017}; from archaic times, cultures have been intrigued and inspired  by the Galaxy -- in classical Latin {\it Via Lactea}\footnote{Dictionaries and online sources: Cambridge, at \url{https://dictionary.cambridge.org/dictionary/}. Oxford, at \url{https://www.oxfordlearnersdictionaries.com/}. Merriam-Webster, at \url{https://www.merriam-webster.com/}. Online Etymology Dictionary, at \url{https://www.etymonline.com/} 
Stanford Encyclopedia of Philosophy, at \url{https://plato.stanford.edu/}. Ancient History Encyclopedia, at \url{https://www.ancient.eu/}.} -- the hazy band that can be visible across a cloudless and unpolluted night sky (Figure~\ref{fig1}).
The ancient philosophers speculated about what was that luminous band, the Platonist philosopher  Plutarch described it as a cloudy circle, from Greek {\it galaxias kyklos}, literally milky circle ({\it k\'yklos} \lq\lq wheel\rq\rq). Across the ages, arts and sciences have portrayed the origin and meaning of the Galaxy; today we know that it is the Milky Way seen from inside (Figure~\ref{fig2}a); from the spiral arm where the Solar System is located, the so-called Orion Arm or Orion Spur  \citep{Bok1950,Bok1950MNSSA}. The Milky Way is only one galaxy among hundreds of billions, a neighbour galaxy is  Andromeda (Figure~\ref{fig2}b)  \citep{Conselice2016,Urquhart2018,Boardman2020}.

\lq Galaxy\rq \ and \ \lq atom\rq \ are examples of words with Greek etymological roots. They are broadly used in  contemporary science but their meanings evolve with time. Various astronomical objects, chemical elements and molecules also gained their names from Greek. The discovery of the noble gas helium ({\it h\={e}lios} \lq\lq sun\rq\rq) constituted a remarkable example in the recent history of astronomy, physics and chemistry \citep{Kragh2009}. Today we know that hydrogen ({\it hydr-}, hydor \lq\lq water\rq\rq) and helium 
are the lightest and most abundant elements in the Universe, they burn in stars forming heavier elements through nuclear reactions. As part of the life cycle of stars, the chemical elements ejected by a dying star will enrich the cloud in which will take place the birth of the next stellar generation; thus stars are also seen as fossils that preserve the history of their host galaxies \citep{Desilva2015,Kobayashi2016}.

As a big question about the Universe, \citet{Stavinschi2011} also asked: why is nature comprehensible to humans? How is cosmos related to humanity?. Under the perspective of the {\it homo sapiens} evolution, various insights can be found since the earliest interpretations about the Galaxy. Etymologically, Galaxy alludes to the liquid {\it  milk}, the essential food for young mammals. In retrospective, such association is not minor if one considers Charles Darwin's legacy (e.g. \citealt{Darwin1859}), since the physiological synthesis of milk and lactation period are  aspects that continuously drive the mammals' evolution, whose origins date back 200 million years ago \citep{Capuco2009,Thomas2016}.

According to one of the best known stories of the Greek mythology, the Galaxy was formed by the milk spilled from the breast of a goddess, when the child Heracles, the famous hero, son of Zeus (Jupiter) and Alcmene, was pushed from Hera's breast \citep{Bertola2009}. Thus, the whiteness of the milk, secreted from the breast of a nursing mother, was associated with that of the {\it Milky Way}.  Translated from Greek,  ancient texts recorded some thoughts about the Galaxy; in Plutarch's Moral,\footnote{Online source: Online Library of Liberty, at \url{https://oll.libertyfund.org/}.   Plutarch’s Morals. Translated from the Greek by Several Hands. Corrected and Revised by William W. Goodwin, with an Introduction by Ralph Waldo Emerson. 5 Volumes. (Boston: Little, Brown, and Co., 1878). Vol. 3. Chapter I.: Of the Galaxy, or the Milky Way. The text is in the public domain.} Parmenides realised it as a mixture of a thick and thin substance whose colour resembles milk; Anaxagoras as a shadow produced by the relative movement of the Sun and Earth;  Democritus saw it as a splendour which arisen from the coalition of many small bodies \citep{Wintemberg1908}. 

\begin{figure}[t!]
\begin{center}
\includegraphics[width=17cm]{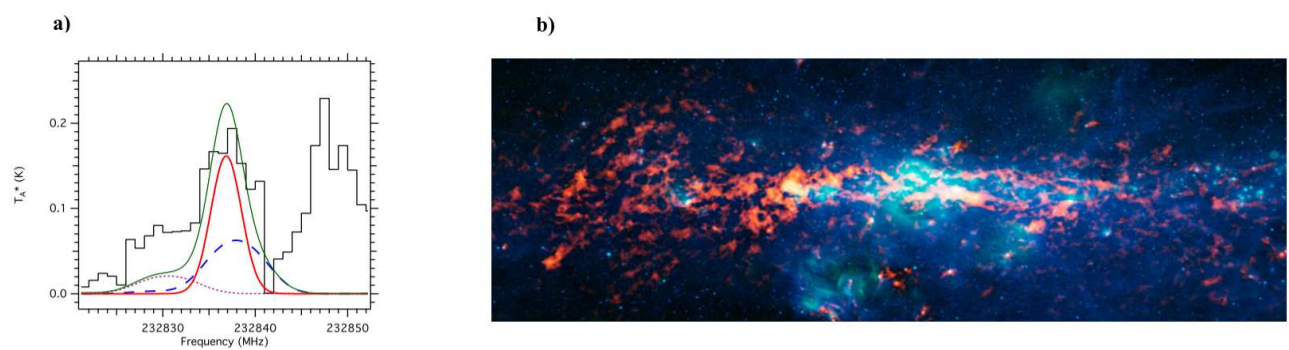}
\end{center}
\caption{a) Result of the search for urea in Sagittarius B2 (N-LMH), the green line represents a total simulated spectrum including both urea (red line) and methyl formate (dashed blue and dotted magenta lines) ({From \citealt{Remijan2014}}). b) Colour-composite image of the Galactic centre and Sgr~B2. Credit: ESO/APEX \& MSX/IPAC/NASA.}\label{fig3}
\end{figure}

Since the renaissance, modern science revoked ideas such as the geocentrism and vitalism, whose  origins date back indeed to older ages. The laboratory synthesis of urea (NH$_2$)$_2$CO, a molecule found in mammals' urine, recently discovered in the Interstellar Medium (ISM) (Figure~\ref{fig3}), see \citet{Remijan2014}, established a new paradigm in the so-called organic chemistry. At the beginning of the 19th century, the chemist Friedrich W\"ohler wrote to his mentor J\"ons Jakob Berzelius: {\it \lq\lq I must tell  you  that  I  can prepare urea without requiring a kidney of an animal, either man or dog}\rq\rq \ \citep{Yeh2007}. For the time being, it was not well known how to transform inorganic substances into organic compounds, so that the production of urea from an aqueous solution of ammonium cyanate, without requiring a \lq vital force\rq,
set a milestone for  chemical synthesis, so that not all the carbon compounds derived from living organisms \citep{Forster2006,Sumiya2019}.

\subsection{The organic side of the Galaxy}

Hydrogen, carbon, nitrogen, oxygen, sulphur and phosphorus are essential elements in biomolecules; they are frequently referred to with the acronym CHNOPS (or CHONPS). The chemical evolution of the biogenic elements, from their cosmic origin to their inclusion in living organisms, constitutes a major topic in astrophysics, astrochemistry and astrobiology \citep{Whittet1993,Nomoto2013,Pizzarello2017}. 

By the fact that they are tangible objects, meteorites have been intensively studied since the
19th century, a period in which was recorded a high number of meteorite falls and fireballs; then, chemists, astronomers, geologists and meteorologists were among the specialists that initially characterised meteorites \citep{Romig1966}. At the present-day, it is known that aminoacids, nucleobases and phosphate are present in meteorites: in a recent study, \citet{Furukawa2019} identified sugar-related compounds in samples of carbonaceous chondrites. Their results also include evidences for  ribose, which is a building block of genetic molecules (e.g.  RNA). The presence of organic molecules in comets is also relevant, since they contain and carry materials of the primitive solar nebula. Using remote observations at radio wavelengths, \citet{Biver2019} describe the identification of organic molecules, such as acetaldehyde (CH$_3$CHO), formamide (NH$_2$CHO) and methyl-formate (HCOOCH$_3$), whose abundances are usually reported with respect to e.g. water or methanol. Those are aspects that unveiled an extraordinary chemistry in space. 

Organic molecules have been detected in different astronomical environments, from objects ranging in size from comets up to external galaxies \citep{Tielens2013}. Glycine, the simplest amino acid,  has been searched for in the ISM but its detection has been controversial \citep{Kuan2003,Snyder2005};  the ROSINA\footnote{ROSINA  - Rosetta Orbiter Spectrometer for Ion and Neutral Analysis. The Mass spectrometer for the Rosetta Mission. Online resource at \url{https://www.esa.int/Science\_Exploration/Space\_Science/Rosetta/ROSINA}.} mass spectrometer provided a more robust result but in the coma of the 67P/Churyumov-Gerasimenko comet, where volatile glycine was confirmed \citep{Altwegg2016,Hadraoui2019}. In the case of the molecule urea, as was mentioned above, it was detected in the Murchison meteorite \citep{Hayatsu1975} and has been searched for towards the high mass star forming region Sgr~B2, Figure~\ref{fig3} (e.g. \citealt{Remijan2014,Belloche2019,Belloche2020}); the interstellar synthesis of urea has also been computationally studied \citep{Slate2020}. 
 
\begin{figure}[t!]
\begin{center}
\includegraphics[width=17cm]{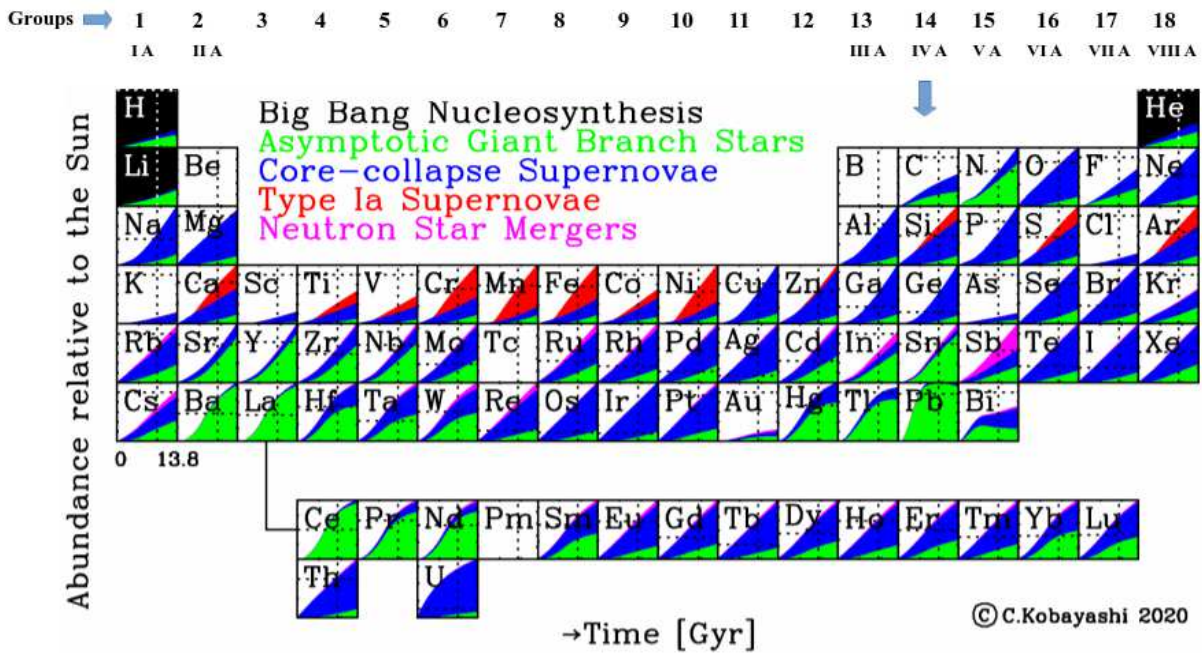}
\end{center}
\caption{Periodic table indicating the cosmic origin and evolution of the chemical elements (From \citealt{Kobayashi2020}); the group numbers were added in the upper part of the periodic table, they are related with the atomic properties. Credit: Chiaki Kobayashi 2020.}\label{fig4}
\end{figure}

 Why do organic compounds matter? Because they are based on the chemistry of carbon, whose most abundant isotope has an atomic and mass number of 6 and 12 ($Z$=6, $A$=12), respectively. As it is known by the nature of triple $\alpha$ reactions, elements with $A \geq$ 12 are produced by stars, the origin of C is associated with low- and intermediate-mass AGB stars \citep{Karakas2010}. Regarding the number of electrons, in the periodic table, carbon occupies a reference cell: it is the first element of the IV A group, in the middle of the I A and VIII~A, which are headed by  hydrogen and helium (Figure~\ref{fig4}), so that carbon has a suitable electronic configuration to form multiple and stable covalent bonds. As a consequence, the element is present in the vast majority of (natural and synthetic) molecules known \citep{Friedman2012,Wencel2013}. In the ISM, carbon can be abundantly found as gas CO, or in condensed phase forming Polycyclic Aromatic Hydrocarbons (PAHs) \citep{Candian2018}. C-bearing molecules are key in chemistry and astrochemistry: from CO, a cosmic diatomic molecule, to the deoxyribonucleic acid (DNA), which can have of the order of 10$^8$ carbon atoms, the search and detection of organic molecules in space  contribute to our understanding of the emergence of life. Research on  molecules like HNCO, H$_2$CO and NH$_2$CHO have shed light about the interstellar chemistry of pre-biotic molecules and species carrying various of the biogenic elements (e.g. \citealt{Saladino2006,Bisschop2007,Mendoza2014,LopezSepulcre2015,Allen2020,Jorgensen2020}). 

The history around the concept of Galaxy is rich; nowadays, astronomy, physics and chemistry converge in specialised areas to study the Milky Way and other Galaxies. In the next sections, we discuss relevant aspects for the subject: astrochemistry and galactic dynamics (Section~\ref{sec2}); state-of-the art in methods and techniques (Section~\ref{sec3});  a discussion considering different astronomical contexts and objects of study (Section~\ref{sec4}); perspectives for future facilities and final remarks (Section~\ref{sec5}). 

\section{Astrochemistry and Galactic dynamics}\label{sec2}

The essential subject matter of astrochemistry, also known as molecular astrophysics, is the study of the formation, destruction and excitation of chemical species in astronomical environments and their influence on the structure, dynamics and evolution of astronomical objects in galaxies \citep{Dalgarno2008,vanDishoeck2018}. To date, more than 200 molecules have been identified in the interstellar and circumstellar media;\footnote{In line with the Cologne Database for Molecular Spectroscopy, CDMS, we refrain from stating an exact number of molecules as detected, since that status usually generates controversies and divergences in the community. More information at \url{https://cdms.astro.uni-koeln.de/classic/molecules}.} among which, there are hydrocarbons, aromatic, inorganic, organic and pre-biotic species \citep{McCarthy2001,Tielens2013,McGuire2018}. The detection of the most abundant molecules, such as CO and its isotopologues (e.g. $^{13}$CO and C$^{18}$O), is used to diagnose the properties and the distribution of molecular clouds across the Galaxy, which is crucial for our understanding concerning Galactic dynamics (e.g. \citealt{heyer15}).

What are the connections between Astrochemistry and Galactic Dynamics, the topic of this collection? Usually, by Galactic dynamics we understand the Stellar Dynamics, basically the study of the orbits of stars. This field of research had impressive progress with the recent release of Gaia results (e.g. Gaia Early Data Release 3, \citealt{GaiaCollaboration2020}) with precise determination of distances, proper motion, photometry and velocity of more than 1 billion  stars. From this data we obtained a much better description of the mass distribution, gravitational forces and Galactic resonances that are acting on the stars, at least within a radius of 2~kpc around the Sun. We must not forget, however, that Galactic Dynamics also includes hydrodynamics. On a large scale point of view, the forces acting on the gas are almost  the same as those acting on the stars, not only because the mass fraction of the Galaxy  in form of gas is small compared to the stellar mass (about 5\%,  \citealt{JamesLequeux2005}) but the mass of the gas is already taken into account in the stellar dynamics, like for instance,  in the gravitational potential derived from the rotation curve. 
The exact nature of the spiral arms is a question of major importance for astrochemistry. In the past, the spiral arms were considered to be large-scale gas shock waves \citep{Roberts1969}, but the stellar mass content has been matter of investigations and continuous revisions over the past decades \citep{Lin1964,Kalnajs1973,Shu2016}. The high resolution HI Nearby Galaxy Survey, THINGS \citep{Walter2008}, revealed the  HI gas is strongly concentrated in the spiral arms, which are narrow structures. The fact that, normally, the arms are thin can also be seen in the image of Andromeda (Figure~\ref{fig2}b). This reinforces the view that the arms are like grooves in the gravitational potential, and that the gas flows along these grooves \citep{Barros2021,Monteiro2021}. The grooves, or elongated potential wells, are due to the larger density of stars in the arm, as the stellar orbits become close to each other,  as in \citet{Kalnajs1973} models, and in the model of our Galaxy  proposed by \citet{Junqueira2013} --see their  Figure~8. 
The spiral arms are known to be the star-formation machines of the Galaxy, as spiral arms tracers are the young and massive stars, like OB-stars, young open clusters, molecular  masers associated with massive stars. The OB stars that we observe in the arms of external galaxies are believed to be the tip of the iceberg of clusters of stars that are not visible to us \citep{Wright2020}. \citet{Falceta2015} presented a detailed  hydrodynamical simulation of interstellar clouds penetrating in the groove-shaped arms, due to the relative velocity of the galactic material with respect to the arms, showing the generation of turbulence and conditions favourable to star formation. Interestingly, at the end of their paper, \citet{Falceta2015} make a comparison with another context of star formation in spiral arms, the context of transient arms \citep{Baba2013}.
They conclude that despite the dynamical differences between the two scenarios, the driving mechanisms of turbulence, which are at the origin of star formation, occur similarly.

The molecular content of the cold gas in spiral arms was observed by \citet{Greaves1996}; they identified various chemical species such as HCO$^+$, HCN, HNC, CN, C$_2$H, C$_3$H$_2$, CS, SiO, N$_2$H$^+$, CH$_3$OH and SO, by means of a technique of looking at absorption lines in front of a radio continuum source, which permits to see the coldest components.

The formation of stars represents in some way the life of our Galaxy, the means by which the Galaxy evolves, and, in particular, the chemical composition of the Galaxy evolves. When we go down in size scales to the star formation cores in the molecular clouds, then hydrodynamics is not governed anymore by the general  potential of the Galaxy, but by very local effects.
Observation of molecules have two main roles. Since stars form in the interior of molecular clouds in regions obscured by the ISM, the molecular emission is a tool (together with mm and sub-mm dust emission) that can reveal what is happening inside of them. Therefore,  astrochemistry is a powerful tool that allows us to determine densities, temperature, and observe gas flows, so that molecules are not only a tool, but also main actors. Although the elements are only synthesised in the interior of stars or in outer layers of stars, in the case of Supernovae, the ISM contributes to transport the elements, in the form of atoms or in the form of molecules or of dust grains, to regions that are distant from the stellar birthplace, and distribute them more uniformly, contributing to the formation of smooth metallicity gradients.

\begin{figure}[h!]
\begin{center}
\includegraphics[width=10cm]{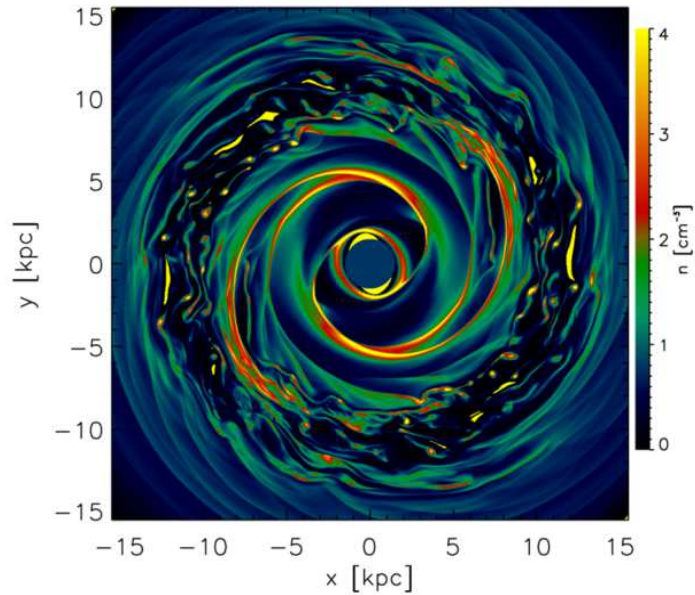}
\end{center}
\caption{Simulation of the density distribution of the galactic gaseous disk when is perturbed by the PERLAS spiral arm potential (From \citealt{Villegas2015}).} \label{fig7-villegas}
\end{figure}

A major feature of the Galactic structure, related to metallicity gradients, is worth to mention here. There is a gap
in the distribution of the ISM, in the form of a ring void of gas, at the corotation radius. \citet{Mishurov2009} called it the \ \lq\lq Cassini-like\rq\rq \ gap. It was observed in the HI surveys of the Galactic disk \citep{Amores2009}, and it appears clearly in the hydrodynamic simulations of the Galactic disk by \citet{Villegas2015} (Figure~\ref{fig7-villegas}). \citet{Monteiro2021} also presents a discussion of the gap, including a figure with a toy model to explain its formation. 

The fact that the gap is larger than 1 kpc in the radial direction, and very deep (the gas density is almost zero in the gap) has enormous consequences.  There is no communication between the gas situated in the two parts, inside and outside the corotation radius. So, the disk of the Galaxy is divided into two parts that have independent chemical evolution. The outer part has a lower metallicity, since the star formation rate is smaller there, due to the smaller gas density, on the average. The transition between the metallicity of the two halves of the galactic disk cannot be called a \lq\lq gradient\rq\rq; \ it is so  abrupt across to the Cassini-like gap, that we prefer to call it the \lq\lq step\rq\rq \ in the metallicity. This step was first discovered by \citet{Twarog1997} who presented a plot of [Fe/H] as a function of galactocentric distances of a sample of Open Clusters, in which we can already  see all the important features:
a step of $-$0.3 dex at 10 kpc (their adopted R0), a gap in the cluster density at 10 kpc (clusters cannot be born in the Cassini-like gap where there is no gas),
and a flat gradient in the outer half-galaxy. The step in [Fe/H] was further studied and confirmed by \citet{Lepine2011}. The existence of this step is an evidence for the long-lived nature of the spiral arm structure. The step height is a consequence of the different metallicity enrichment rate on the two halves of the Galaxy. If the corotation radius were frequently changing its position according to the intermittent arms models, the metallicity step would not have grown. \citet{Lepine2011} estimated a lower limit of 3 Gyrs for the present spiral arm structure.There are other examples of steps in metallicity at corotation, see for instance the step of the [O/H] abundance in M83 seen by \citet{Bresolin2009}; in a similar way, it is an argument in favour of long-lived arms in M83. The flat (stellar) gradient of metallicity in the outer part of the Galactic disk can be understood as an effect of the gas flow in spiral arms, remembering also that stars can only born in arms.
Another important consequence of this bimodal metallicity in Galactic disk is that it may explain the distribution of O-rich and C-rich AGB stars in the Galaxy (a further revision is given in Sec~\ref{sec4p3}). At this point it seems to be important to distinguish the high mass AGB stars with cold circumstellar envelopes (CSE) from the more common low mass AGB stars \citep{Omont1993}. The and O-rich and C-rich AGBs with cold CSE are more clearly separated in the Galaxy.
There is a belief, at least for the more massive stars, that if they are born in an ISM with low metallicity, they will become, in the AGB phase,  C-rich stars, and if the birthplaces have high metallicity, then the stars will become O-rich in the AGB phase (e.g. \citealt{Noguchi2004,Ishihara2011}). One example supporting this concept of the dependence on ambient metallicity is a deep search for OH/IR stars (O-rich AGB stars with cold CSE) that was performed by \citet{Goldman2018} in the Small Magellanic Cloud (SMC), a low metallicity galaxy. They made long integration on a dozen of luminous, long-period, large-amplitude variable stars and no one confirmed to be an O-rich star.  We will come back  to the discussion of AGB stars in a later section.

The metallicity step is a feature of the ISM  which affects young stars. However, there is no obstacle  for  stars, as soon as formed, to start crossing the corotation gap, since it is a gap of hydrodynamical origin, and the stars have very little interaction with the gas. It is possible to find older stars that were born in the low metallicity region that are presently in the high metallicity regions and vice-versa \citep{Lepine2014}.
The radial migrations turn it more difficult to observe the metallicity step,  depending on the sample of stars used; it is better observed using young stars. For instance, the gap at 8 kpc and the difference in metallicity of the inner and outer disk can be noticed in the sample of HII regions (\citealt{Paladini2004}, see their Figure~4).
The authors do not mention the observed feature, because this was not understood at that time. Interestingly, none of the major models of chemical evolution of the Galaxy mention the important points that we discussed here. 
Concerning the importance of the corotation resonance  being close to the Sun, see also the implications in the solar neighbourhood regarding the Hercules stream, discussed in a later section.

\section{Approaches and methodologies}\label{sec3}

\begin{figure}[h!]
\begin{center}
\includegraphics[width=16cm]{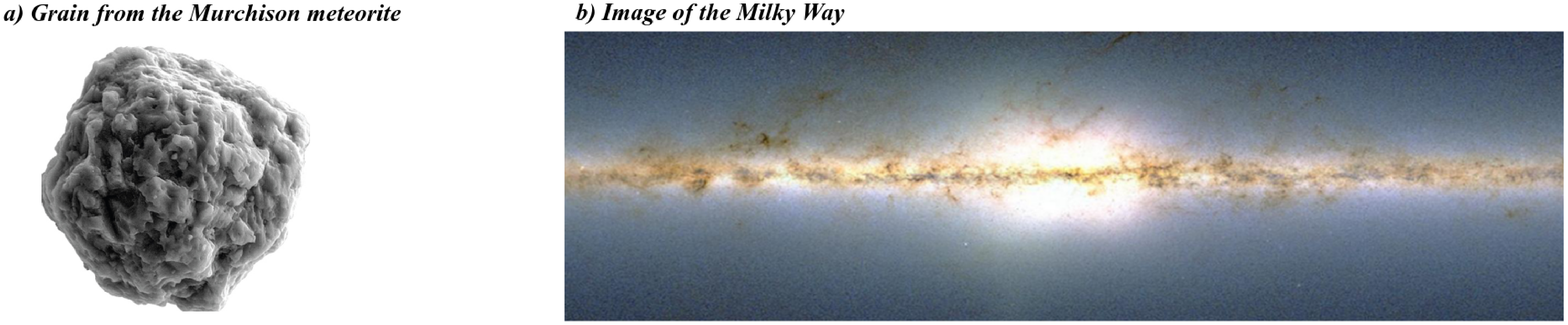}
\end{center}
\caption{The universe as can be explored from a micro to a cosmic scale: a) Scanning electron microscope image showing the morphology of a presolar silicon carbide (SiC) grain extracted from the Murchison CM2 meteorite (From \citealt{Heck2020}); b) Infrared map of the Milky Way exhibiting the disk, as a plane, and bulge of the Galaxy. APOGEE-2 Background (https://www.sdss.org/surveys/apogee-2/apogee-2-background/).}\label{micro-to-macro}
\end{figure}

In view of the relevance that astrochemistry has from a micro to a cosmic scale (Figure~\ref{micro-to-macro}), in this section we focus on how it integrates observational, experimental and theoretical  approaches (Figure~\ref{approaches-astrochemistry}) to  study the physical and chemical conditions of the  interstellar gas and dust in Galaxies. 

\subsection{Molecules in space: observations from infrared to radio wavelengths}

\begin{figure}[t!]
\begin{center}
\includegraphics[width=12cm]{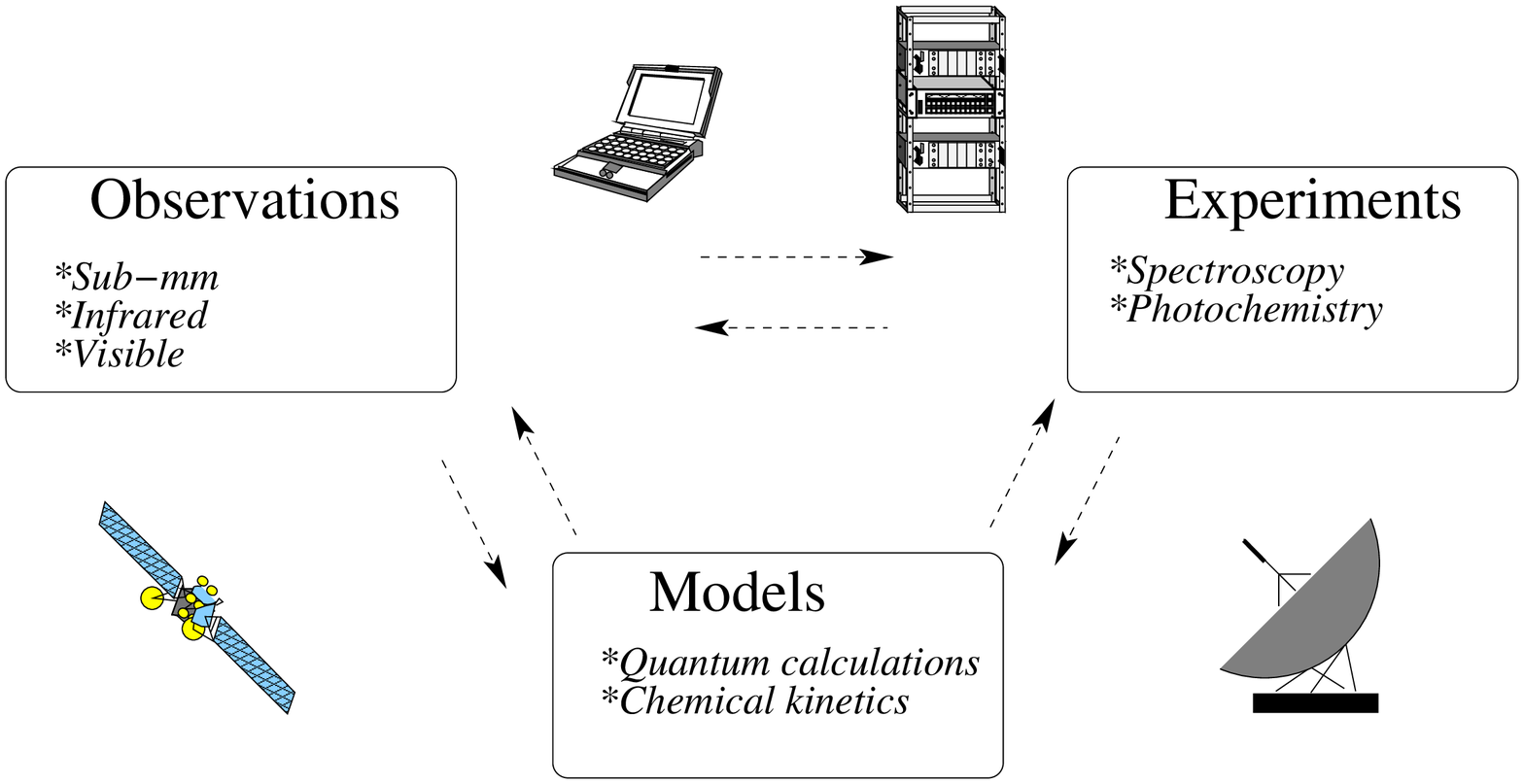}
\end{center}
\caption{Approaches in astrochemistry: observations, models and experiments are carried out to detect, analyse, mimic and explain physico-chemical processes in astronomical environments.}\label{approaches-astrochemistry}
\end{figure}

Molecules emit radiation from transitions which are denoted by electronic, vibrational and rotational quantum levels, thus a total energy is defined as $E$ = $E_{electronic}$ + $E_{vibrational}$ + $E_{rotational}$. The electronic transitions are typically the most energetic and can stimulate photodissociation processes, such as those that take place in the upper atmosphere (stratosphere and above). Electronic transitions have frequencies ranged from the visible to the UV region of the spectrum; vibrational transitions are typically observed in the infrared range; rotational transitions, which are the weakest in energy, are approximately ranged from the sub-mm to cm range of the electromagnetic spectrum \citep{Cook2003,Wilson2013}. 

Astrochemistry has been benefited in the past decades with the construction of telescopes to observe interstellar molecules \citep{vanDishoeck2018}. Since the early 1960s, with the advent of radio astronomy, molecules have been detected with ground-based facilities like the NRAO 36ft, IRAM 30m, Nobeyama 45m and GBT 100m. Galactic sources like Sgr-B2, IRC+10216, TMC-1 and Orion record a vast number of molecular detections \citep{McCarthy2001,McGuire2018}. From infrared to radio wavelengths, latter facilities like the {\it Spitzer} Space Telescope, {\it Herschel} Space Observatory, the Stratospheric Observatory for Infrared Astronomy (SOFIA), and the (sub)millimeter telescopes in the Atacama desert, such as the Atacama Pathfinder Experiment (APEX), the Atacama Submillimeter Telescope Experiment (ASTE), and the Atacama Large Millimeter/submillimeter Array (ALMA) have contributed to the state-of-the art of observational astrochemistry. In addition to molecular surveys based on abundant species (e.g. CO and CH$_3$OH), those new facilities have provided results about rare species, such as ArH$^+$ \citep{Schilke2014}, HeH$^+$ \citep{Gusten2019,Neufeld2020}, p-H$_2$D$^+$ \citep{Brunken2014} and complex organic molecules like formamide and N-methylformamide (NH$_2$CHO and CH$_3$NHCHO); the latter studied by \citet{Belloche2017} combining experiments, modelling and observations with ALMA towards Sgr B2(N2). 

The interstellar molecules are generally listed by their number of atoms. In the particular case of the so-called complex organic molecules (COMs), with a different connotation e.g. in organic chemistry or biochemistry, it stands for interstellar molecules containing an organic functional group and usually more than 5 atoms \citep{Herbst2009}, such as the C$_2$H$_4$O$_2$ isomers methyl formate (HCOOCH$_3$), glycolaldehyde (CH$_2$OHCHO) and the acetic acid (CH$_3$COOH), which have been detected in Sgr B2(N) \citep{Xue2019}. Examples of lists of interstellar and circumstellar molecules can also be found at
{\it The Astrochymist}\footnote{The Astrochymist: \url{http://www.astrochymist.org/astrochymist\_ism.html}.} (D. Woon), in which species are ordered by first year of reported discovery, also including the astronomical sources, wavelength and bibliographic information.

Along with the  advances about detection of molecules, efficient codes have also been developed to infer the physical and chemical conditions from observations.
As part of the data analysis, Local Thermodynamic Equilibrium (LTE) hypotheses are  generally evaluated  \citep{Goldsmith1999,Roueff2020}; however, collisional rate coefficients and non-LTE calculations are also important to estimate e.g. gas kinetic temperatures and volume densities.
 Public programs like RADEX,\footnote{RADEX: \url{https://home.strw.leidenuniv.nl/~moldata/radex.html}.} which is a one-dimensional non-LTE radiative transfer code \citep{vanderTak2007},  permit the estimation of physical conditions from a given observational dataset. The RADEX code is also an alternative tool to the population diagram method (e.g. \citealt{Goldsmith1999}), which relies upon the availability of various optically thin emission lines. Collisional rate coefficients are essential for statistical equilibrium calculations. The Leiden Atomic and Molecular Database (LAMDA)\footnote{LAMDA: \url{https://home.strw.leidenuniv.nl/~moldata/}.} contains spectroscopic and collisional rate coefficients for a number of astrophysically interesting chemical species, including data for 4 atomic/ionic species and 37 molecules \citep{Schoier2005}. A new revision of LAMDA, including its current status, recent updates, and future plans, is described in \citet{vanderTak2020}, where planned updates consider data for noble gas species, extensions of existing collisional data to more transitions, additional collisional partners and data for molecular isotopologues. This conjunction of theoretical, experimental and observational methods (Figure~\ref{approaches-astrochemistry})
represents an example about how astrochemistry is developed nowadays.

\subsection{Experimental astrochemistry: Laboratory work}

Atomic and molecular spectroscopy, gas-phase and ice chemistry are essential subjects to understand the physics and chemistry of the ISM, they are part of the laboratory work that benefit astrochemistry. Regarding spectroscopy, the association of an astronomical spectral signature to a given molecule depends on the availability of rest frequencies and spectroscopic properties which are usually obtained from laboratory data and quantum chemical calculations. The Cologne Database for Molecular Spectroscopy (CDMS) is an exceptional resource founded in 1998 that provides an updated catalogue of molecular species  \citep{end2016}; various of them have been detected and reconfirmed in astronomical sources usually observed in radio wavelengths, however, in particular cases, the catalogue contains entries for transitions in the far- and mid-infrared, such as those for species as C$_3$, C$_3$O$_2$, CH$^+$, C$_2$H, HCO$^+$ and N$_2$H$^+$. The Splatalogue database for astronomical spectroscopy compiles resources such as CDMS, the JPL molecular spectroscopy catalogue \citep{pic1998} and the Lovas/NIST list of recommended rest frequencies for observed interstellar molecular microwave transitions \citep{Lovas2004}.

Although interstellar molecules are generally observed in gaseous phase, it is well known that the solid state plays an important role in the synthesis of molecules \citep{Boogert2015,Linnartz2015}.  As part of the lifecycle of gas and dust during the stellar evolution, the cold conditions in dark interstellar clouds favour a rich ice chemistry in which  molecules are produced in icy dust grains. The Infrared Space Observatory and the {\it Spitzer} provided important evidences on the composition of interstellar ices \citep{Gibb2004,Oberg2011};  in particular, as part of the {\it Spitzer} ice legacy, \citet{Oberg2011} estimated that the median ice composition H$_2$O:CO:CO$_2$:CH$_3$OH:NH$_3$:CH$_4$:XCN (e.g. OCN$^-$) is 100:29:29:3:5:5:0.3 and 100:13:13:4:5:2:0.6 towards low- and high mass protostars, respectively. The ice chemistry is relevant to explain aspects like the formation of complex organic molecules, e.g. ethylene glycol (H$_2$C(OH)CH$_2$OH) \citep{Fedoseev2015}, and isotopic exchange reactions, e.g. in H$_2$O, H$_2$CO ice analogues \citep{Ratajczak2009,Oba2012}, since gas chemical reactions alone cannot explain the presence and abundance of various interstellar molecules.

The study of interstellar ice analogues has also provided clues about how life arose in our planet. {\it Chirality} is a property that some molecules exhibit, living organisms use left-handed (L) aminoacids in proteins and right-handed (D) sugars in RNA and DNA; the appearance of molecular chirality in meteorites and/or comets might provide clues on the asymmetric evolution of life on Earth \citep{Kondepudi1990,Evans2012}. In the ISM, \citet{McGuire2016} reported the discovery of the chiral molecule propylene oxide (CH$_3$CHCH$_2$O) towards the Sagittarius B2 North molecular cloud. Experiments have demonstrated that ices irradiated with UV circularly polarised light can yield a given stereo-specific photo-chemistry, in that way \citet{Modica2014} found an enantiomeric excess during the synthesis of amino acids in ice analogues irradiated at two different photon energies. Among the 16 amino acids identified in the experiment, they found an L-enantiomeric excess in five of them. Based on the known biochemical processes, the search and study for chiral asymmetry is discussed as a potential bio-signature in the solar system \citep{Goesmann2017,Glavin2020}.

\subsection{Calculations and modelling in astrochemistry}

The different conditions and inaccessibility to interstellar regions demand elaborating models to understand the physical and chemical processes in space. Astrochemical models can be used to study the composition and chemical reactions of the ISM, they use to include gas-solid reactions depending on the evolutionary stage of the interstellar regions. Reactions of astrochemical interest have been experimentally and theoretically investigated by research groups worldwide. The Kinetic Database for Astrochemistry (KIDA)\footnote{KIDA: \url{http://kida.astrophy.u-bordeaux.fr/}.} and the UMIST database for Astrochemistry (UDfA)\footnote{UMIST RATE12: \url{http://udfa.ajmarkwick.net/index.php}.} are web-based resources providing lists of species and chemical reactions between them, such databases compile rate coefficients from the literature and are continuously updated.  \citet{McElroy2013} presented one of the latest release of the UDfA, which consists in a new reaction network (labelled Rate12) containing 6173 gas-phase reactions involving 467 chemical species. Among the novelties, various anion chemical reactions, deuterium exchange reactions and surface binding energies for neutral species are available. \citet{Wakelam2015} presents the 2014 KIDA network for interstellar chemistry (labelled kida.uva.2014), which includes a total of 7509 reactions involving 489 species. In comparison with previous versions (kida.uva.2011), 446 rate coefficients changed and 1038 new reactions were added. Comparisons between both databases have been discussed,  \citet{McElroy2013} compares Rate12 with the kida.uva.2011 considering a dark cloud model. For instance, among 62 modelled species, considering observations of TMC-1, 38 agree at a time of 2.5 $\times$ 10$^5$~yr. They also discuss differences for O-bearing molecules like acetaldehyde (CH$_3$CHO) and formic acid (HCOOH) \citep{Hamberg2010,Vigren2010}, which are associated with newly measured dissociative recombination rate coefficients for reactions with electrons. 

KIDA also provides astrochemical codes,\footnote{KIDA Codes: \url{http://kida.astrophy.u-bordeaux.fr/codes.html}.} such as the Nautilus gas-grain code whose latest version allows to simulate the chemical evolution of the ISM considering three phases, i.e., gas, grain surface and grain mantles \citep{Iqbal2018}; the gas chemistry uses the kida.uva.2014 network, the grain chemical network is that  presented in  \citet{Garrod2007} and \citet{Ruaud2015}. Such type of codes and chemical models  find important applications in sources like cold cores, which in conjunction with  sub-mm observations, might explain the desorption and formation of COMs in icy mantles  \citep{Vasyunin2017}. 

\section{The objects of study: a large and special issue}\label{sec4}

In order to provide an overview about astronomical objects connecting Astrochemistry and Galactic Dynamics, in this section we present a discussion from three different perspectives: namely, $i.$ the chemistry of the early Universe, and external galaxies; $ii.$ star-forming regions and $iii.$ evolved stellar objects such as AGB stars. Such topics are also seen in perspective for future investigations integrating the methodologies described in \S~\ref{sec3}.   
\subsection{Chemistry of the early Universe}

After the primordial nucleosynthesis, a chemistry started in the recombination era when the ions produced in the Big Bang recombined with free electrons. The first molecules and molecular ions were simple species, such as H$_2$, HD, HeH$^+$ and LiH. Given the difficulties to observe them, the chemistry of those species is mainly inferred from models \citep{Galli2013}. In the case of the helium hydride ion (HeH$^+$),   Table~\ref{tab:primordial-chemistry} lists  chemical equations of formation and destruction reported in the literature. 

\citet{Miller1992} presented and discussed evidences for H$_3^+$ and HeH$^+$ in the envelope of the supernova 1987A. \citet{Zinchenko2011} searched for HeH$^+$ and CH in a high-redshift quasi-stellar object. In a recent work, \citet{Gusten2019} identified HeH$^+$ in the planetary nebula NGC 7027 using data from the SOFIA airborne observatory; they identified the pure rotational HeH$^+$ $J$=1--0 ground-state transition at 149.137~$\mu$m. \citet{Neufeld2020} confirmed the discovery of HeH$^+$ in NGC 7027;they identified rovibrational lines at 3.51629 and 3.60776~$\mu$m using the NASA Infrared Telescope Facility (IRTF) on Mauna Kea.

\begin{table*}[!t]
\centering
\caption{Primordial chemistry of HeH$^+$.  Examples of chemical reactions of formation and destruction of HeH$^+$. }
\label{tab:primordial-chemistry}
\begin{tabular}{lll}
\hline  
Reactions: formation & Type & Reference    \\
\hline   
 H$_2^+$ + He $\longrightarrow$ HeH$^+$ + H	&	Ion-neutral	&	\citet{Theard1974}\\
 H$^+$ + He $\longrightarrow$ HeH$^+$ + $h\nu$	&	Radiative association	&	\citet{Zygelman1998} \\
\hline
Reactions: destruction & Type & Reference    \\
\hline
 HeH$^+$ + e$^- \longrightarrow$ He + H	&	Dissociative recombination	&  \citet{Novotny2019}	\\
 H + HeH$^+ \longrightarrow$ He + H$_2^+$	&	Ion-neutral	& \citet{Karpas1979}	\\
\hline
\end{tabular}
\end{table*}

Heavier chemical elements would start to appear after the formation of the first stars, or Population III stars. Galactic chemical evolution models are important to understand the baryonic matter behaviour in galaxies \citep{1997nceg.book.....P,2012ceg..book.....M}. As a type of feedback mechanism, molecules like CO and SiO provide C and Si to produce dust grains, e.g. in the form of amorphous carbon and/or refractory silicate compounds concurrently. The grain surfaces are essential to catalyse the H$_2$ synthesis, since the gas phase routes do not account for the H$_2$ formation and abundance. \citet{Marassi2015} quantified the role of core-collapse supernovae as the first cosmic dust polluters, they carried out models about dust formation considering grain species like amorphous carbon (C), Al$_2$O$_3$, Fe$_3$O$_4$, MgSiO$_3$, Mg$_2$SiO$_4$ and SiO$_2$, in the gas phase they considered reactions involving CO and SiO. Depletion processes are important in the dust formation;  grains can grow at high accretion rates by condensation of gas molecules like CO and SiO.  In the context of Type II supernovae at redshift $z \geq 5$, \citet{Schneider2004} found that the grain condensation starts $\sim$~150 days after the explosion, when the temperature of the expanding gas falls from $T\thickapprox$10$^4$~K to $T\thickapprox$500~K, yielding dust masses in the 0.01--10 $M_{\odot}$ range. Therefore, in such events dust grows and acts as an effective cooler agent thermalising the gas phase.

Currently, there are no observational constraints on the
Pop III SNe events.
However, observational results point for evidences of large amounts of dust at high-redshifts. The reddening of background quasars and Ly$\alpha$ systems at high redshift ($z \textgreater$ 6) require large column densities of dust to be explained \citep{2009ApJ...703..642C}. Detection of dust thermal emission appears from high-redshift QSOs from the Sloan Digital Sky Survey (SDSS) out to redshift 6.4. From the IR luminosities of the hyperluminous galaxy SDSS J1148+5251 at redshift $z$ = 6.4 \citep{2003A&A...406L..55B}, the amount of dust required to explain the results would be about 2$\times$10$^8M_{\odot}$ \citep{Schneider2004,2009ApJ...703..642C}. These results cannot be accounted for by processes efficient for dust formation in the local Universe such as the winds of AGB and supergiant stars and, the colliding winds of Wolf–Rayet stars or the ejecta of core-collapse supernovae, mainly due to their long evolutionary timescales of around a few Gyr. Therefore, the short timescales for supermassive primordial stars evolution offer the best explanation for the high amounts of dust found at high redshifts \citep{Schneider2004,2008ApJ...683L.123C,2009ApJ...703..642C}.

With observatories like ALMA, new evidences demonstrate the important role of dust in the evolution of galaxies. \citet{Dunlop2017} conducted a deep ALMA imaging in the Hubble Ultra Deep Field ($z \thickapprox$2.15), they estimated that $\sim$~85 \% of the total star formation is hidden in dust.

\subsubsection{\underline{Astrochemistry in neighbouring and more remote galaxies}}

The first observations of molecules in external galaxies were reported in the 1970s; \citet{Weliachew1971} recorded the first detection of a molecular line outside the Milky Way, which was OH identified in the NGC 253 and M82 galaxies. From the abundant H$_2$ \citep{Thompson1978} to pre-biotic species like NH$_2$CHO \citep{Muller2013}, various of the molecules observed in the Milky Way have also been found in external galaxies \citep{Viti2016}.

In the vicinity, the Large Magellanic Cloud (LMC), located at about 50 kpc, is a galaxy that is seen nearly face-on with an inclination angle of 35$^\circ$. The metallicity in the LMC is Z$\sim$0.5 Z$_{\odot}$ \citep{keller06}, and the gas-to-dust ratio is a factor of 4 higher than in the Milky Way. This galaxy has many active
star-forming and HII regions (e.g. \citealt{paron16,och17}). Therefore, the LMC is a unique laboratory for studying the effects of metallicity on molecular gas and star formation at high spatial resolution in an extragalactic medium. Indeed, several complex molecules have been detected towards hot cores in the LMC. \citet{Sewilo2018} reported the first extragalactic detection of the COMs dimethyl ether (CH$_3$OCH$_3$) and methyl formate (HCOOCH$_3$); along with CH$_3$OH, a chemical precursor, those species were detected toward the N 113 star-forming region in the LMC. Molecular studies like those can shed light on chemical processes in the past universe, where
the metallicity was significantly lower than the present-day galaxies (see also \citealt{shimo20}).

\subsection{Astrochemistry in star forming regions}

Interstellar molecules  can be found in different conditions and environments, either in nearby objects of our solar system or distant galaxies. The study of interstellar molecules has helped to better understand not only  the evolutionary stages of stellar formation, but also the physical and chemical properties of Galactic nebulae and star forming regions, which are known to be one of the main tracers of the spiral structure of the Galaxy \citep{Reid2019}. The molecular spectra may contribute to the determination of important properties like temperature,  density, abundance, as well as kinematical properties (expanding or collapsing motions, rotations, etc.) of the molecular clouds in which the molecules  are embedded. Although  the high-mass star formation is not well understood, it is believed to occur inside giant molecular clouds; the resulting stages in the formation of high-mass young stellar objects (YSOs) can be summarised as {\it a)}  Infrared dark clouds (IRDCs), which are massive and cold ($T \sim$ 10 K) clouds where the collapse takes place, {\it b)} warm/hot ($T \sim$ 100--300 K) inner envelopes heated by the central protostar with sizes up to  $\sim$ 0.1 pc and densities of 10$^{7-8}$ cm$^{-3}$, known as hot cores (e.g. \citealt{Kurtz2000,Tan2014,Motte2018,Beaklini2020}). This stage is characteristic to be strong emitter of rotational lines of complex molecular species. In fact, the detection of COMs, which are present in dense and hot gas, is the strongest indicator of the hot core phase and its the most distinctive feature. At last, {\it c)} the ultracompact HII region, powered by the newly formed YSO. For the case of low-mass star forming regions the phenomena and stages may  include {\it a)} molecular outflows, {\it b)} hot corinos, which are the low mass version of hot cores, and {\it c)} protoplanetary disks (e.g. \citealt{Ceccarelli2007,Walsh2016,Lefloch2018,Belloche2020}). High mass protostellar objects also exhibit a complex chemistry, studies have been dedicated to detect and model the presence of molecules (e.g. CH$_3$OH, H$_2$CO, CH$_3$CN, HNCO and NH$_2$CHO) in YSOs, disks and outflows \citep{Isokoski2013,Choudhury2015}.  
     
Hot molecular cores and hot corinos are believed to be objects preceding the formation of high-mass and low-mass YSOs, respectively. They are both characterised for being strong emitters of high energy rotational lines of rare molecular species  at  mm and sub-mm   wavelengths. They will determine the chemical richness of star forming regions in the Galaxy which,  as previously mentioned, are the main tracers of the spiral structure of our Galaxy. Then, it is crucial to detect and study complex organic molecules in hot cores and hot corinos which  allows a better understanding of chemical process  that are taking place on them and to better comprehend the chemical compositions of the spiral structure of the Galaxy. In the last few years, several detections of complex molecules were obtained towards  a small number of Galactic hot cores/corinos  (paradigmatic cases are  Sgr B2, Orion-KL, W51, NGC 6334I).  For these reasons, it is crucial to increase the number of detections of complex  molecules in high and low-mass  star forming regions during the latest phases of star formation.

\subsection{Evolved stars and circumstellar envelopes}
\label{sec4p3}

\begin{figure}[h!]
\begin{center}
\includegraphics[width=5cm]{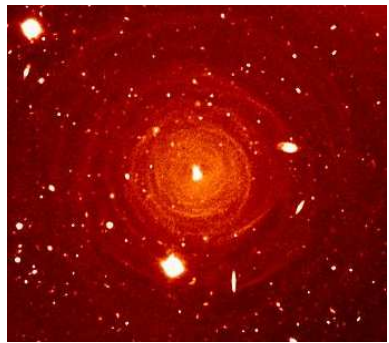}
\end{center}
\caption{Image of IRC+10216 and its shells. The source is a near C-rich AGB star, various chemical species including anions (e.g. CN$^-$) have been detected in its circumstellar envelope \citep{Agundez2010}. Image credits: Izan Leao (Universidade Federal do Rio Grande do Norte, Brazil).}\label{irc101216}
\end{figure}

The AGB stars and their circumstellar envelopes exhibit a rich chemistry in heavy elements, ions (cations and anions), radicals and molecules; various of the chemical species detected in space have been discovered in circumstellar shells \citep{Xiaohu2016}. For example,  IRC+10216 (Figure~\ref{irc101216}) is one of the nearest and most investigated AGB stars; observational and theoretical studies have unveiled aspects on its chemistry, structure and evolution \citep{leao2006,Cernicharo2015}. AGB stars are classified according to their carbon-to-oxygen abundance ratios; on the one hand, they are denoted as carbon-rich (C-rich) AGB stars if C/O $>$ 1; on the other hand, oxygen-rich (O-rich) AGB stars if C/O $<$ 1; as an intermediate  category, the S-type AGBs stand for those sources with C/O $\thickapprox$ 1 values \citep{Brunner2018}. 

\begin{figure}[h!]
\begin{center}
\includegraphics[width=17cm]{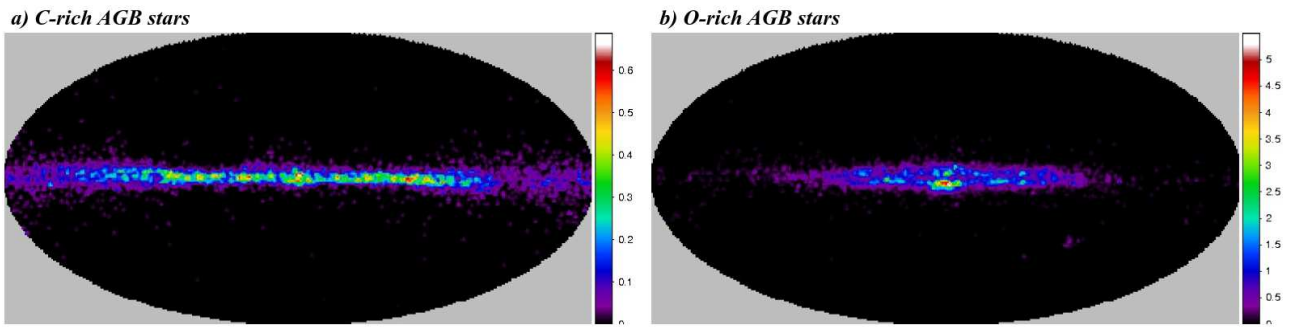}
\end{center}
\caption{Distributions of C- and O-rich AGB samples (From: \citealt{Ishihara2011}).}\label{AGBs-Galaxy}
\end{figure}

In the Milky Way, O-rich AGB stars exhibit silicate dominated circumstellar features. C-rich AGB stars exhibit signatures associated with hydrocarbons and carbonaceous dust. Regarding their Galactic distributions, O-rich AGBs are more concentrated towards the Galactic centre while C-rich ones are distributed more uniformly across the Galaxy \citep{LeBertre2003}. \citet{Ishihara2011} confirmed such trends in a  study of C- and O-rich AGB stars using a mid-infrared all-sky survey (Figure~\ref{AGBs-Galaxy}), also proposing that the metallicity gradient with galactocentric distance might be one of the possible factors to explain the spatial distributions of C-rich and O-rich AGB stars; a comprehensive study covering the entire Milky Way is still needed to understand the dependence on the metallicity and Galactic environment. Evolved stars are also important in studies about the chemical enrichment of the ISM in galaxies, that can be specially studied in the Local Group of galaxies where AGB stars are useful to study the structure, chemical enrichment and galaxy evolution \citep{Javadi2019}. Surveys such as the {\em Herschel} inventory of the Agents of Galaxy Evolution (HERITAGE) \citep{Meixner2010} have allowed to investigate the presence of evolved stars candidates in the LMC and SMC with metallicites of 0.5 and 0.2$Z_{\odot}$, respectively \citep{Russell1992,Jones2015}. Studies highlight the importance of using the ratios between C- and O-rich AGB stars to evaluate the metallicity distribution in galaxies like the Magellanic Clouds, unveiling that stars in the SMC are in general older and metal poorer than the LMC, in line with the fact that the SMC is considered as a more primitive and less evolved galaxy than the LMC \citep{Cioni2003,Cioni2006}. We  remind here the discussion presented in Section~\ref{sec2} on the two halves of the Galactic disk
that are separated by an abrupt step in metallicity. The abrupt separation is 
attenuated for  populations of  not too young stars, like AGBs, since part of them might have suffered migration across the corotation gap, which is a barrier only for the gas.

We will now focus on the C-rich AGB stars with thick circumstellar dust envelopes, since these AGBs or post-AGBs have the potential to have strong influence on the chemical evolution of the ISM. These stars have  quite high mass loss rates, frequently over $10^{ -5}  M_{\hbox{$\odot$}}$ per year. \citet{Epchtein1990} made a crude estimate of the number of C-rich AGB stars with thick circumstellar dust envelopes in the Galaxy and of their mass loss rates. They reached the conclusion that the total mass returned to the Galaxy is 0.5 $M_{\hbox{$\odot$}}$ per year; as a rough estimation, in 2 Gyr this would amount to $10^{9}  M_{\hbox{$\odot$}}$, which corresponds to 20\% of the mass of the ISM estimated by \citet{JamesLequeux2005}. One may wonder, therefore, if part of the molecules that are found in the cold ISM  did not originate in the winds of AGB stars. 

IRC+10216 is considered as the prototype of the C-rich carbon stars with thick envelopes. \citet{Matthews2015} discovered the presence of a shell of neutral hydrogen around IRC+10216, with an estimated diameter of 0.8 pc.The authors did their observations in the 21 cm HI line, but based on simple calculations of mass-loss rates they conclude  that the bulk of the stellar wind is in molecular form, not only in the form of HI. The number of published works on IRC+10216 is enormous. There are results like the Plateau de Bure and ALMA high resolution observations, the observations of dust shells, the discovery of  a dust lane and the binary nature of the star. We will only focus on the molecular species that have been discovered. A list of the molecules detected in IRC+ 10216 according to \citet{Ziurys2006} is given in Table~\ref{tab:molecules-AGBs}, as well as the molecules reviewed by \citet{McGuire2018}. Recently, \citet{Cernicharo2019} discovered the magnesium-bearing species MgC$_3$N and MgC$_4$H in the same source. 

\begin{table*}[!t]
\centering
\caption{List of  various of the molecular identifications carried out in IRC+10216.}
\label{tab:molecules-AGBs}
\setlength{\tabcolsep}{24pt}
\begin{tabular}{llllll}
    \hline
    \multicolumn{6}{c}{Chemical species in IRC+10216}\\
\hline
CO         &  C$_2$H      & HC$_3$N    & C$_2$S       & SiO     & NaCl \\
CS         &  C$_3$H      & HC$_5$N    & C$_3$S       & SiS     & AlCl \\
CN         &  C$_3$O      & HC$_7$N    & C$_5$S       & SiC     & KCl  \\
CN$^-$     &  C$_4$H      & HC$_9$N    & C$_2$N       & SiN     & AlF  \\
HCN        &  C$_5$H      & H$_2$C$_4$ & C$_3$N       & SiC$_2$ & MgCN \\
C$_2$H$_2$ &  C$_6$H      & H$_2$C$_6$ & C$_3$N$^-$   & SiC$_3$ & MgNC \\
HNC        &  C$_6$H$^-$  & HC$_2$N    & C$_5$N       & SiCN    & MgC$_3$N \\
C$_2$H$_4$ &  C$_7$H      & NC$_2$P    & C$_5$N$^-$   & SiNC    & MgC$_4$H \\
CH$_4$     &  C$_8$H      & CH$_3$SiH$_3$ & HC$_4$N      & SiC$_4$ & FeCN \\
NH$_3$     &  C$_8$H$^-$  & C$_3$      & c-C$_3$H$_2$ & SiH$_4$ & AlNC \\
H$_2$S     &  C$_2$       & C$_5$      & CH$_3$CN     &         & KCN     \\
           &              &            & CP           &         & NaCN     \\
           &              &            & C$_2$P       &         &      \\
           &              &            & PN           &         &      \\
           &              &            & HCP          &         &      \\
\hline
\end{tabular}

Note: The list was organised considering e.g.  \citet{Ziurys2006}, \citet{Glassgold1996} \citet{Agundez2014}, the 2018 Census of molecular detections presented by \citet{McGuire2018}, and current detections reported by \citet{Cernicharo2019}.
\end{table*}

The fact that many molecules are composed of elements heavier than O, such as 
Mg, Si, Na, S, Al, P, Fe, shows that indeed the AGB stars are contributing to the chemical evolution of the Galaxy. 
\citet{Nyman1993} compared the molecular abundances of IRAS 15194-5144, another very bright C-rich AGB star with those of IRC+10216. They performed a survey in the 1.3 and 3 mm bands with the
Swedish-ESO Submillimetre Telescope (SEST), and detected 29 transitions of 14 species and some of their isotopologues.
\citet{Woods2003} observed a rich and complex chemistry towards a sample of carbon rich circumstellar envelopes observed in both the northern and southern skies, for which they used the Onsala Space Observatory 20m and SEST antenna, respectively. They identified more than 20 chemical species,  e.g. CN, C$_2$H, HC$_3$N and CH$_3$CN, whose chemistry in the outher envelope can be explained through photochemical models.
Almost all the species detected in IRC+10216 were also present in IRAS 15194-5144 and there is a good correlation between the abundances. Similarly, \citet{Zhang2009} performed a line survey of the carbon AGB  star CIT6 and found that the spectral properties of CIT6 are consistent with those of IRAS+10216. We are allowed, therefore, to consider that the set of molecules in Table~\ref{tab:molecules-AGBs}, at least for the strongest lines, is characteristic of C-rich AGB stars.
We now come to the question of the observations of \citet{Greaves1996}, who made a survey of molecules in \lq\lq spiral arms clouds\rq\rq \ and found HCO$^+$, HCN, HNC, CN, C$_2$H, C$_3$H$_2$, CS, SiO, N$_2$H$^+$, CH$_3$OH and SO, which are basically the same species expelled by C-rich AGB stars, except that now that there are two new oxygen-rich molecules, which could be explained if the  line-of-sight  crosses regions  containing also O-rich AGB stars. 

In a recent work, \citet{Cristallo2020} claims that the majority (90\%) of presolar SiC grains found in primitive meteorites (e.g. Figure~\ref{micro-to-macro}a) originate from ancient AGB stars, whose ejected mass were merged into the Solar System during its formation. Thus, the role of AGB stars in chemical enrichment is significant for producing $^{12, 13}$C, $^{14}$N, F, $^{25, 26}$Mg, $^{17}$O and slow neutron-capture process (s-process)  \citep{Kobayashi2020}. Evolved stellar objects are relevant for galactic studies, the similar molecular content of AGB-stars mass loss and in spiral arms clouds is not a sufficient proof that the gas clouds in the arms originate partly from AGB star, but it suggests that it is a question that merits to be further investigated. 

\subsection{Dynamics very close to the Sun, moving groups and stellar streams}

It is known that there exists groups of stars situated around and very near  the Sun, that have anomalous velocities, each of these groups forming a kind of \lq\lq wind\rq\rq \ of stars travelling in a given direction. \citet{Antoja2018}  presents an historical perspective and  an updated list of the moving groups. The interest of these groups for astrochemistry is that they offer the opportunity to investigate what was the composition of the gas where the stars were born. Let  us take one example, the Hercules stream, which is well studied. According to an investigation on the origin of moving groups by \citet{Barros2020}, the Hercules stream could be mostly associated with the 8/1 and 12/1 Lindblad resonances, believed to be strong resonances in a 4 arms spiral galaxy. The stellar orbits at these resonances is illustrated in Figure~5 (bottom) of \citet{Michtchenko2018}.   See  Figure~7 of \citet{Barros2020} to follow the discussion on the origin of the stars. 
Starting from their present position, by integrating their orbits to the past (counter-clockwise in that figure), we should find their birthplaces. The orbits present many loops, but they can be considered  to be contained in a kind of torus (a concept much used in dynamics). Since the stars  are expected to born in spiral arms, we should look for the first crossing of the  torus (or of the stellar orbits) with an arm. This is situated on the Sagittarius-Carina arm; we must conclude, therefore, that most stars of the Hercules stream were born in that arm.
It must be remembered that the figure represents the arms in their own  standard of rest, which rotates with a velocity of the order of 28 km/s/kpc, in the inertial frame.
\citet{Ramya2016} performed a detailed study of the stellar population of the Hercules stream. They found an average metallicity of the order of [Fe/H]= 0.15, which is perfectly compatible with the stellar orbits discussed above, which have a radius a little smaller than that of the Sun. They found an average age of the Hercules stars of about 1.0 Gyrs, which again is compatible, in order of magnitude, with the travel time between the probable birthplace and the present-day position. There is a small uncertainty in the exact position of the Sagittarius-Carina arm which is possibly not a perfect spiral. Finally, the surprising velocity of the Hercules stream stars in the LSR frame, with a large component in the galactic radial direction, could be explained by a part of loop situated close to the Sun (see Figure~5, bottom left, of \citealt{Michtchenko2018}).
Note that the ideas presented in this section are not at all a consensus in the literature, see e.g. \citet{Monari2017}, \citet{Asano2020}, among others,  with a great variety of proposed models.

\section{Final remarks and perspectives}\label{sec5}

\begin{figure}[t!]
\begin{center}
\includegraphics[width=7cm]{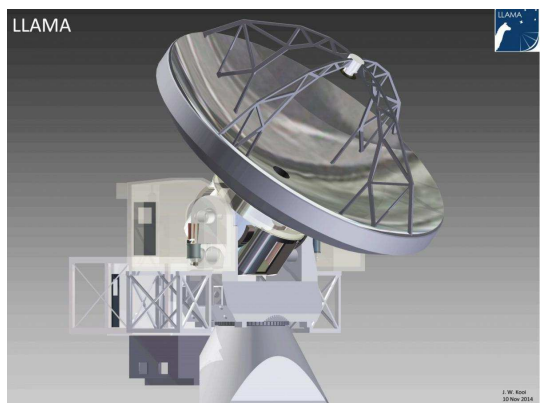}
\end{center}
\caption{Future LLAMA radio telescope.}\label{antena}
\end{figure}

Astrochemistry looks like a flourishing area of research, with such an incredible amount papers published
in a few recent years, bringing new molecules, new ideas, and new paths for future research, that we had  to make choices in our presentation. Possibly our interest in radio astronomy and in particular in the possibilities of the ALMA interferometer had some influence.
Most of the authors of the present paper are involved, directly or indirectly, in the LLAMA Argentinian-Brazilian project to construct of a 12m sub-mm radio telescope in the Andes (Figure~\ref{antena}), at 4800m altitude. The first light will take place in 2022, and the first two receivers will work in  band 5 (163--211 GHz) and band 9 (602--720 GHz) -ALMA Bands. The cooled receivers are ready and waiting in the NOVA labs in Groningen (NL). These frequency bands are very rich in molecular lines, and certainly a good fraction of the observing time will be dedicated to Astrochemistry. This group is happy to ascertain that there will be no lack of ideas for competitive use of LLAMA. The association of Astrochemistry with Galactic Dynamics will be an additional source of original ideas and important contributions to science.

\section*{Conflict of Interest Statement}

We declare that the research was conducted in the absence of any commercial or financial relationships that could be construed as a potential conflict of interest.

\section*{Author Contributions}

Authors contributed with the content and revision of the manuscript.

\section*{Acknowledgments}

Professor Dr. Jacques R. D. L\'epine contributed with insightful discussions, we are grateful for his comments and suggestions to connect the topics discussed in this manuscript.

\bibliographystyle{frontiersinSCNS_ENG_HUMS} 
\bibliography{frontiers}

\end{document}